\documentclass[prb,showpacs,floatfix,preprint,aps]{revtex4}
\usepackage{epsfig,amsmath,amssymb}
\newcommand{\be}{\begin{equation}}
\newcommand{\ee}{\end{equation}}

\newcommand{\bit}{\begin{itemize}}
\newcommand{\eit}{\end{itemize}}
\newcommand{\bea}{\begin{eqnarray}}
\newcommand{\eea}{\end{eqnarray}}
\newcommand{\stateX}{\(\sqrt{3}\)\(\times\)\(\sqrt{3}\)}

\newcommand{\Kagome}{kagom\'e}
\newcommand{\kagome}{{\Kagome} }
\begin{document}
\title
{
Absence of long-range order in a spin-half Heisenberg antiferromagnet on the
stacked kagom\'{e} lattice
}
\author
{
D. Schmalfu{\ss}$^{a}$, J. Richter$^{a}$ and D. Ihle$^{b}$ 
}
\affiliation
{
$^{a}$Institut f\"ur Theoretische Physik, Otto-von-Guericke Universit\"at
Magdeburg, 39016 Magdeburg, Germany \\   
}
\affiliation
{
$^{b}$Institut f\"ur Theoretische Physik,Universit\"at Leipzig, 04109 Leipzig,
Germany\\
}                     

\begin{abstract}
We study the ground state of a spin-half Heisenberg antiferromagnet on
the stacked kagom{\'e} lattice by
using a spin-rotation-invariant Green's-function method. Since the pure
two-dimensional kagom\'{e} antiferromagnet is most likely a magnetically
disordered quantum
spin liquid,  we investigate the question whether the coupling of kagom\'{e}
layers
in a stacked three-dimensional system
may lead to a magnetically ordered ground state. 
We present spin-spin correlation functions and 
correlation lengths. For comparison we apply also
linear spin wave theory. 
Our results provide strong evidence that the system remains
short-range ordered independent of the sign and the strength of the interlayer
coupling. 
\end{abstract}


\maketitle

\section{Introduction} \label{intro}
In frustrated quantum spin lattices the interplay of quantum and
frustration effects causes interesting physics, see e.g. the recent reviews
[\onlinecite{lhuillier01,moessner01,lhuillier03,wir04}].
Special attention was focused on the problem of the ground state
(GS)
nature of the Heisenberg antiferromagnet (HAFM) on the 
two-dimensional (2D) kagom\'{e} lattice. 
In the classical limit the GS of the HAFM on the 
kagom\'{e} lattice exhibits a huge non-trivial
degeneracy (see, e.g. Refs. \onlinecite{moessner01,harris92,chalker92,huse92}). 	
Intensive work over the last decade on the 
spin half quantum version of the model  
\cite{harris92,zeng,singh,chalk,chub,sach,leung,yang,asa,sind,
lecheminant97,waldtmann98,mila98,sindzingre00,farnell01,
yu,bern,jump,male02,senthil03}
led to the conclusion  
that for the quantum HAFM the GS is  most likely a
short-range ordered spin liquid without any kind of long-range order (LRO). 
However, we mention that the 
GS properties in the quantum case are far from being 
fully understood, yet. Most of the conclusions are drawn from
finite-lattice results of up to $N=36$ sites.
Very recently, the rotation-invariant
Green's-function method introduced by Kondo and Yamaji\cite{kondo} has been
successfully applied on  
the spin-half HAFM on the 2D
kagom\'{e} lattice\cite{yu,bern}. 
In the framework
of this approach one obtains also a spin-liquid GS with an energy per spin
$E_{0}/JN=-0.4296$ which agrees well with the best available results 
obtained by other methods like the exact diagonalization  
\cite{wir04,zeng,lecheminant97,waldtmann98} 
($E_{0}/JN=-0.4344$) and the coupled-cluster method  \cite{farnell01}
($E_{0}/JN=-0.4252$).
Note that additional second-neighbor
couplings may lead to a magnetically ordered GS phase
\cite{lecheminant97,shankar00,narumi04}.  

It is well-known that the dimensionality is crucial for the existence of
order.  
Roughly
spoken one can say, the higher the dimensionality of the spin system
the more the influence of quantum or thermal fluctuations is
suppressed.
For the GS of quantum spin systems one finds several examples, where
the physical properties change basically  going from one to
two dimensions or from two to three dimensions. For instance, the 
transition from the linear-chain to the  square-lattice HAFM was studied in
Refs. \onlinecite{affleck,mutter,ihle1,derzhko}.           
Another example is the frustrated $J_1$-$J_2$ HAFM, which exhibits a
magnetically disordered quantum
paramagnetic  GS phase around $J_2 \sim 0.5J_1$ for the 2D square lattice (see, e.g.
Refs. \onlinecite{doucot88,schulz92,ri93,sorella00,SIH01,sushkov01,capriotti01} 
and references therein) 
but does not show this kind of quantum phase for the 3D body-centered cubic
lattice \cite{schmidt,oitmaa04}.
Moreover, we know from the Mermin-Wagner theorem \cite{merm}
that the interlayer
coupling in quasi-2D Heisenberg magnets is crucial for 
magnetic ordering at finite
temperatures. The role of the interlayer coupling in quasi-2D systems
becomes particularly interesting 
if the decoupled layers themselves are magnetically
disordered, e.g.  due to strong frustration. Though this question is relevant for real 3D solids with a
layered arrangement of magnetic atoms it
has been less studied yet. 

In this paper we address this question for the HAFM on a congruently 
stacked \kagome
lattice. For such a system it is not known, 
whether an appropriate interlayer coupling may lead to a 
semi-classically ordered magnetic GS. 
However, one can expect that in the
special limit of
infinite ferromagnetic interlayer coupling the physics of the classical 
$O(3)$ \kagome antiferromagnet  is obtained (see Sec. \ref{ferro}).
We emphasize, that the Lanczos exact
diagonalization of finite lattices being the most powerful method 
to study the HAFM in the pure 2D \kagome case is inappropriate in 3D. 
Therefore we will 
discuss this question in our paper by 
the above mentioned spin-rotation-invariant Green's-function method, 
having
in mind that this method was successfully applied to the pure 2D case.
In particular, we will calculate 
the correlation functions and the correlation lengths. 
In addition, we will discuss the
linear spin-wave theory (LSWT) for comparison.

\section{The Model}
\begin{figure}
\begin{center}
\epsfig{file=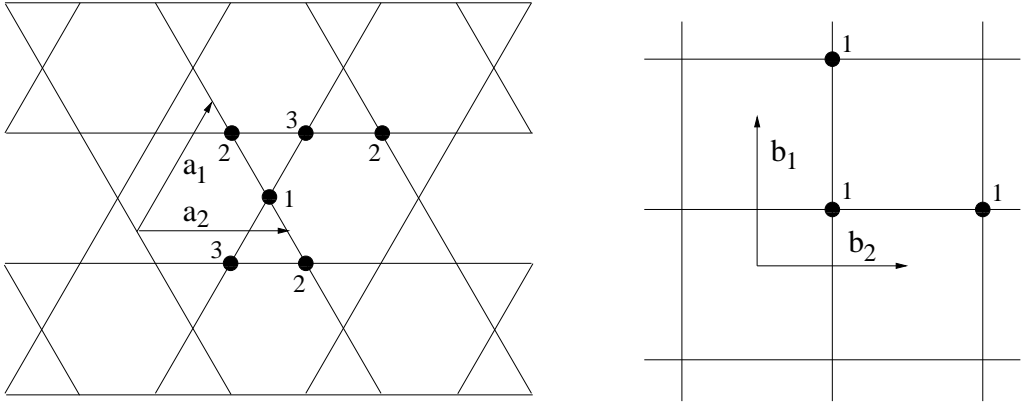,scale=0.50,angle=0.0}
\end{center}
\caption{Sketch of one  kagom\'{e} layer in  the congruently stacked 
lattice with its in-plane geometrical
unit vectors
${\bf{a}}_{1}=\left(0,2,0\right),{\bf{a}}_{2}=\left(\sqrt{3},1,0\right)$.
The out-of plane unit vector 
${\bf{a}}_{3}=\left(0,0,1\right)$ is not shown. 
Within a geometrical unit cell the
spins are distinguished by a running index $\alpha=1,2,3$.}
\label{fig1}
\end{figure}

We consider congruently stacked kagom\'{e} layers shown in Fig.\ref{fig1}.
The vertices of this stacked lattice are occupied by 
$N$ spins one-half interacting 
via the Heisenberg exchange coupling
\begin{equation}
H=\frac
{1}{2}\sum_{m\alpha,n\beta}J_{m\alpha,n\beta}{\bf{S}}_{m\alpha}{\bf{S}}_{n\beta},
\label{eq1}
\end{equation}
where the sum runs over all unit cells (labeled by $m$ and $n$) and 
 all spins within a unit cell (labeled by running indices
$\alpha$ and $\beta$, see Fig.~\ref{fig1}). 
One unit cell
contains three spins, therefore the number of cells is $N/3$. The exchange
coupling $J_{m\alpha,n\beta}$ is
non-zero for nearest neighbors (NN), only. We introduce two different
exchange parameters $J_{\parallel}$ and  $J_{\perp}$ according to
$J_{m\alpha,n\beta}\rightarrow
J_{\parallel}\left(1-\delta_{\alpha,\beta}\right)+J_{\perp}\delta_{\alpha,\beta}$.
Since we are interested in the  stacked \kagome HAFM, we consider
$J_{\parallel}>0$ but allow both signs for $J_{\perp}$.  

\section{Classical ground state and linear spin wave theory}
\label{lswt}
We start with a brief discussion of 
the classical ground state (GS) of (\ref{eq1}). Because there is no additional
frustration caused by $J_{\perp}$, the classical GS within a certain 
layer  is not
modified, and the nontrivial huge degeneracy due to corner-sharing
triangles \cite{moessner01,harris92,chalker92,huse92} 
is not lifted. The only effect of a
ferromagnetic (antiferromagnetic) 
interlayer coupling $J_{\perp}$ is the parallel (antiparallel)  orientation
of interacting spins of adjacent layers. Thus for a certain 
classical GS the spin configurations in all layers   are
identical.  
Two particular variants for the classical GS, namely the
so-called \stateX $\mbox{}$ state and the so-called
$q=0$ state (see e.g.\ Fig.\ 17 in Ref.\ \onlinecite{wir04}),
are often used for discussing possible magnetic 
LRO in the \kagome
lattice. These two particular planar states can also be considered 
as variants of possible GS ordering for the stacked \kagome lattice.

Let us remind the reader of the fact, 
that starting from the \stateX $\mbox{}$ state as well as from the 
$q=0$ state 
 the linear spin-wave theory (LSWT) 
for the pure 2D \kagome HAFM \cite{harris92,asa}
leads to one flat
zero mode and two degenerate modes producing 
divergent integrals in the sublattice magnetization, which might be
 taken as a
hint on the absence of semi-classical magnetic order.

For the stacked (3D) system we start also from both the 
\stateX $\mbox{}$ state and  the 
$q=0$ state and perform 
the LSWT as usual. Taking into account
that we have three spins per unit cell we have to introduce three different
kinds of magnons. As in the 2D case\cite{harris92,asa} the spin wave 
spectrum is equivalent for all coplanar 
 classical GS configurations because the directional
cosine of in-plane neighbors is always $-1/2$ while the directional cosine
of out-of-plane neighbors is either $1$ or $-1$ depending on the sign of
$J_{\perp}$. For $J_{\perp}>0$ we obtain for the spin-wave dispersions
\begin{eqnarray}
\omega_{1{\bf{q}}}&=&s\sqrt{4J^{2}_{\perp}\left(1-\cos^{2}
q_{z}\right)+6J_{\parallel}J_{\perp}\left(1+\cos
q_{z}\right)},\label{eq2a}\\
\omega_{2{\bf{q}}}&=&s\sqrt{a_{{\bf{q}}}+b_{{\bf{q}}}},\quad\omega_{3{\bf{q}}}=s\sqrt{a_{{\bf{q}}}-b_{{\bf{q}}}}
\label{eq2}
\end{eqnarray}
with
\begin{eqnarray}
a_{{\bf{q}}}&=&4J^{2}_{\parallel}\left(2\cos^{2}\frac
{q_{y}}{2}+\left(\cos^{2}\frac {\sqrt{3}q_{x}}{2}+\cos^{2}\frac
{q_{y}}{2}\right)\left(1-2\cos^{2}\frac
{q_{y}}{2}\right)\right)\nonumber \\
&+&4J^{2}_{\perp}\left(1-\cos^{2}
q_{z}\right)+3J_{\parallel}J_{\perp}\left(3-\cos q_{z}\right),\\
b_{{\bf{q}}}&=&J_{\parallel}J_{\perp}\left(1-3\cos
q_{z}\right) D_{\bf q}, \\
D_{\bf q} &=&
\sqrt{9+16\cos^{4}\frac
{q_{y}}{2}+16\cos^{2}\frac {\sqrt{3}q_{x}}{2}\cos^{2}\frac
{q_{y}}{2}-8\cos^{2}\frac
{\sqrt{3}q_{x}}{2}-24\cos^{2}\frac {q_{y}}{2}} \label{dq}.
\end{eqnarray}
For $J_{\perp}<0$ we find
\begin{eqnarray}
\omega_{1{\bf{q}}}&=&s\sqrt{4J^{2}_{\perp}\left(1-\cos
q_{z}\right)^{2}-6J_{\parallel}J_{\perp}\left(1-\cos
q_{z}\right)},\label{eq3a}\\
\omega_{2{\bf{q}}}&=&s\sqrt{a_{{\bf{q}}}+b_{{\bf{q}}}},\quad\omega_{3{\bf{q}}}=s\sqrt{a_{{\bf{q}}}-b_{{\bf{q}}}}
\label{eq3}
\end{eqnarray}
with
\begin{eqnarray}
a_{{\bf{q}}}&=&4J^{2}_{\parallel}\left(2\cos^{2}\frac
{q_{y}}{2}+\left(\cos^{2}\frac {\sqrt{3}q_{x}}{2}+\cos^{2}\frac
{q_{y}}{2}\right)\left(1-2\cos^{2}\frac
{q_{y}}{2}\right)\right)\nonumber \\
&+&4J^{2}_{\perp}\left(1-\cos
q_{z}\right)^{2}-9J_{\parallel}J_{\perp}\left(1-\cos q_{z}\right),\\
b_{{\bf{q}}}&=&J_{\parallel}J_{\perp}\left(1-\cos
q_{z}\right) D_{\bf q}.
\end{eqnarray}
The quantity  $s$ in Eqs. 
(\ref{eq2a}), (\ref{eq2}), (\ref{eq3a}), (\ref{eq3})
is the spin quantum number which can be considered
as parameter of the model (\ref{eq1}) within the LSWT.  
If $J_{\perp}$ goes to zero, the known flat zero mode as well as the both
degenerate modes\cite{harris92,asa} are recovered.
But even for non-zero $J_{\perp}$ we have a flat zero mode 
$\omega_{1{\bf{q}}}$ and degenerate $\omega_{2{\bf{q}}}$ and $\omega_{3{\bf{q}}}$
in the  $q_x-q_y$ plane for certain $q_{z}$.
For $J_{\perp}>0$, $\omega_{1{\bf{q}}}$ is a flat zero mode 
at $q_{z}=\pi$,
and $\omega_{2{\bf{q}}}$ and $\omega_{3{\bf{q}}}$ become 
degenerate at any $q_{x},q_{y}$ for $q_{z}=\arccos 1/3$.
For $J_{\perp}<0$, $\omega_{1{\bf{q}}}$ becomes a flat zero mode in 
the $q_x-q_y$ plane for
$q_{z}=0$, and $\omega_{2{\bf{q}}}$ and $\omega_{3{\bf{q}}}$ are degenerate at
any $q_{x},q_{y}$ also for $q_{z}=0$. 
Thus, in both cases there is no finite sublattice magnetization 
due to the resulting logarithmic divergencies in the involved integrals.
We conclude, that the LSWT yields very similar behavior for the pure 2D
and the 3D stacked case.

For completeness we give the expression for the GS
energy per spin  $E_0/N$:
\begin{equation}
\frac {E_0}{N}=
-\left(J_{\parallel}+\left|J_{\perp}\right|\right)s\left(s+1\right)+\frac
{1}{2N}\sum_{{\bf{q}}}\sum_{m=1}^{3}\omega_{m{\bf{q}}},
\label{eq4}
\end{equation}
where the  $\omega_{m{\bf{q}}}$ are the respective 
spin wave dispersions from Eqs.
(\ref{eq2a}), (\ref{eq2}) or (\ref{eq3a}), (\ref{eq3}).

\section{Spin-rotation-invariant Green's-function theory}
The Green's function method is one of the most powerful techniques for the
investigation of quantum many-body systems and was successfully applied to
spin systems over many decades. The rotation-invariant decoupling scheme
was  introduced  by Kondo and Yamaji \cite{kondo}    
to study the one-dimensional Heisenberg antiferromagnet and ferromagnet 
at finite temperatures.
This decoupling scheme was 
developed to
improve the description of magnetic short-range order and allows to describe
magnetic order-disorder transitions driven by quantum fluctuations as well
as by thermal fluctuations.
Later on the  spin-rotation-invariant
Green's-function theory was used to discuss several two-dimensional models, for example the
pure HAFM\cite{shima,winter}, the  
frustrated $J_1$-$J_2$ HAFM\cite{BB94,SIH01} and the spatially anisotropic HAFM on the square lattice 
\cite{ihle1}
as well as the
doped square-lattice HAFM ($t$-$J$-model) \cite{feng,WI98}.
The  quasi-two-dimensional
and the three-dimensional HAFM have been investigated, too \cite{siu,ihle2}.

Let us  outline the main ideas of this method based on the
equation-of-motion technique. To evaluate the relevant correlation
functions we have to calculate a set of
Fourier-transformed Green's functions
$\left\langle\left\langle S^{+}_{{\bf{q}}\alpha};
S^{-}_{-{\bf{q}}\beta}\right\rangle\right\rangle_{\omega}$ which are related
to the 
dynamic spin susceptibilities by 
$\chi^{+-}_{{\bf{q}}\alpha\beta}\left(\omega\right)=-\left\langle\left\langle
S^{+}_{{\bf{q}}\alpha};S^{-}_{-{\bf{q}}\beta}\right\rangle\right\rangle_{\omega}$. 
Their equation of motion
reads
\begin{equation}
\omega\left\langle\left\langle
S^{+}_{{\bf{q}}\alpha};S^{-}_{-{\bf{q}}\beta}\right\rangle\right\rangle_{\omega}=
\left\langle\left[S^{+}_{{\bf{q}}\alpha},S^{-}_{-{\bf{q}}\beta}\right]_{-}\right\rangle
+\left\langle\left\langle
i\dot{S}^{+}_{{\bf{q}}\alpha};S^{-}_{-{\bf{q}}\beta}\right\rangle\right\rangle_{\omega}.
\label{eq5}
\end{equation}
Supposing rotational symmetry we have  $\left\langle
S^{z}_{m\alpha}\right\rangle=0$ for any spin and,  as a result, 
$\left\langle\left[S^{+}_{{\bf{q}}\alpha},
S^{-}_{-{\bf{q}}\beta}\right]_{-}\right\rangle \equiv 0$.
Furthermore, we have 
$\langle S^{z}_{m\alpha}S^{z}_{n\beta}\rangle = \frac{1}{2}\langle
S^{+}_{m\alpha}S^{-}_{n\beta}\rangle$, i.e. it is sufficient to calculate 
$\langle S^{+}_{m\alpha}S^{-}_{n\beta}\rangle$ to know all components 
of the
correlation function $\langle {\bf S}_{m\alpha}{\bf S}_{n\beta}\rangle$.
Going beyond the RPA decoupling\cite{tya} we consider 
in a second step the equation of motion for 
$\left\langle\left\langle
i\dot{S}^{+}_{{\bf{q}}\alpha};S^{-}_{-{\bf{q}}\beta}\right\rangle\right\rangle_{\omega}$, 
\begin{equation}
\omega\left\langle\left\langle
i\dot{S}^{+}_{{\bf{q}}\alpha};S^{-}_{-{\bf{q}}\beta}\right\rangle\right\rangle_{\omega}
=\left\langle\left[i\dot{S}^{+}_{{\bf{q}}\alpha},S^{-}_{-{\bf{q}}\beta}\right]_{-}\right\rangle
+\left\langle\left\langle
-\ddot{S}^{+}_{{\bf{q}}\alpha};S^{-}_{-{\bf{q}}\beta}\right\rangle\right\rangle_{\omega}.
\label{eq6}
\end{equation}
The combination of Eqs. (\ref{eq5}) and (\ref{eq6})  
yields
\begin{equation}
\omega^{2}\left\langle\left\langle S^{+}_{{\bf{q}}\alpha};S^{-}_{-{\bf{q}}\beta}\right\rangle\right\rangle_{\omega}=\left\langle\left[i\dot{S}^{+}_{{\bf{q}}\alpha},S^{-}_{-{\bf{q}}\beta}\right]_{-}\right\rangle
+\left\langle\left\langle
-\ddot{S}^{+}_{{\bf{q}}\alpha};S^{-}_{-{\bf{q}}\beta}\right\rangle\right\rangle_{\omega}.
\label{eq7}
\end{equation}
The operator $\; -\ddot{S}^{+}_{{\bf{q}}\alpha}=[[S^{+}_{{\bf q}\alpha},H],H]$ 
contains products of three
spin operators along NN sequences. Those operator products
were treated in the spirit of the decoupling scheme by Shimahara and Takada
\cite{shima}. This decoupling is performed in the site representation of the
Green's functions. For example, the operator product 
$S^{-}_{A}S^{+}_{B}S^{+}_{C}$
is replaced by $\eta_{A,B}\left\langle S^{-}_{A}S^{+}_{B}\right\rangle
S^{+}_{C}+\eta_{A,C}\left\langle S^{-}_{A}S^{+}_{C}\right\rangle
S^{+}_{B}$, where $A,B,C$ represent spin sites. The quantities  
$\eta_{\gamma,\mu}$  are vertex parameters introduced to improve the
approximation scheme.
In the
minimal version of the theory we introduce just as many vertex parameters as
independent conditions for them can be formulated. Because there are
 just two
such conditions (see below) we consider two different parameters $\eta_{\parallel}$ and
$\eta_{\perp}$ attached to intralayer and interlayer correlators, respectively.
After the decoupling
we write Eq. (\ref{eq7}) in a compact matrix form 
omitting the running indices $\alpha$ and $\beta$, 
\begin{equation}
\left(\omega^{2} - F_{{\bf{q}}}\right)\chi^{+-}_{{\bf{q}}}\left(\omega\right)=
- M_{{\bf{q}}},
\label{eq8}
\end{equation}
where $F_{{\bf{q}}}$ and $M_{{\bf{q}}}$ are the frequency and momentum
matrices, respectively. Since the unit cell contains three spins, 
$F_{{\bf{q}}}$, $M_{{\bf{q}}}$ and $\chi^{+-}_{{\bf{q}}}$
are $3\times 3$-matrices. For the sake of brevity we do not give the lengthy
expressions for the matrix elements of $F_{{\bf{q}}}$ and $M_{{\bf{q}}}$ but
give their eigenvalues, only.
Both matrices  are hermitean
and commute with each
other. Hence, the solution of (\ref{eq8})  in terms of the common set
of normalized eigenvectors $|j{\bf{q}}\rangle$
of $\chi^{+-}_{{\bf{q}}}$ reads
\begin{equation}
\chi^{+-}_{{\bf{q}}}\left(\omega\right)=-\sum_{j=1}^{3}\frac
{m_{j{\bf{q}}}}{\omega^{2}-\omega^{2}_{j{\bf{q}}}}\left|j{\bf{q}}\right\rangle\left\langle
j{\bf{q}}\right|.
\label{eq9}
\end{equation} 
For the eigenvalues $m_{j{\bf{q}}}$ of $M_{{\bf{q}}}$ and
$\omega^{2}_{j{\bf{q}}}$ of $F_{{\bf{q}}}$ we find
\begin{eqnarray}
m_{1{\bf{q}}}&=&-12J_{\parallel}c_{1,0,0}-4J_{\perp}c_{0,0,1}\left(1-\cos
q_{z}\right), \label{eq10a}\\
\omega^{2}_{1{\bf{q}}}&=&3J^{2}_{\parallel}\left(1+2\eta_{\parallel}\left(2c_{1,0,0}+c_{1,1,0}+c_{2,0,0}\right)\right)\nonumber\\
&&+J^{2}_{\perp}\left(1-\cos q_{z}\right)\left(1+2\eta_{\perp}\left(c_{0,0,1}+c_{0,0,2}\right)
-4\eta_{\perp}c_{0,0,1}\left(1+\cos q_{z}\right)\right)\nonumber\\
&&+4J_{\parallel}J_{\perp}\left(\eta_{\perp}c_{0,0,1}\left(1-\cos
q_{z}\right)-2\eta_{\perp}c_{1,0,1}\left(1+\cos
q_{z}\right)\right.\nonumber\\
&&+\left. 7\eta_{\perp}c_{1,0,1}-3\eta_{\parallel}c_{1,0,0}\cos
q_{z}\right), \label{eq10aa}
\eea
\bea
m_{2{\bf{q}}}&=&-6J_{\parallel}c_{1,0,0}-4J_{\perp}c_{0,0,1}\left(1-\cos
q_{z}\right)
-2J_{\parallel}c_{1,0,0} D_{\bf q}, \\
\omega^{2}_{2{\bf{q}}}&=&J^{2}_{\parallel}\left(\frac
{3}{2}+3\eta_{\parallel}\left(2c_{1,0,0}+c_{1,1,0}+c_{2,0,0}\right)\right.\nonumber\\
&&+8\eta_{\parallel}c_{1,0,0}\left(2\cos^{4}\frac
{q_{y}}{2}+2\cos^{2} \frac {\sqrt{3}q_{x}}{2}\cos^{2}\frac
{q_{y}}{2}\right.
-\left.\left.\cos^{2}\frac {\sqrt{3}q_{x}}{2}-3\cos^{2}\frac
{q_{y}}{2}\right)\right)\nonumber\\
&&+J^{2}_{\perp}\left(1-\cos q_{z}\right)\left(1+2\eta_{\perp}\left(c_{0,0,1}+c_{0,0,2}\right)
-4\eta_{\perp}c_{0,0,1}\left(1+\cos q_{z}\right)\right)\nonumber\\
&&+J_{\parallel}J_{\perp}\left(-8\eta_{\perp}c_{1,0,1}\left(1+\cos
q_{z}\right)-2\eta_{\perp}c_{0,0,1}\left(1-\cos q_{z}\right)\right.\nonumber\\
&&-\left.6\eta_{\parallel}c_{1,0,0}\cos q_{z}+22\eta_{\perp}c_{1,0,1}\right)\nonumber\\
&&+\frac
{J_{\parallel}}{2}\left(J_{\parallel}\left(1+2\eta_{\parallel}\left(2c_{1,0,0}+c_{1,1,0}+c_{2,0,0}\right)\right)\right.\nonumber\\
&&+\left.4J_{\perp}\left(\eta_{\perp}c_{0,0,1}\left(1-\cos
q_{z}\right)+\eta_{\perp}c_{1,0,1}-\eta_{\parallel}c_{1,0,0}\cos
q_{z}\right)\right) D_{\bf q} 
\label{eq10b},
\eea
\bea
m_{3{\bf{q}}}&=&-6J_{\parallel}c_{1,0,0}-4J_{\perp}c_{0,0,1}\left(1-\cos
q_{z}\right) + 2J_{\parallel}c_{1,0,0}D_{\bf q}, \\
\omega^{2}_{3{\bf{q}}}&=&J^{2}_{\parallel}\left(\frac
{3}{2}+3\eta_{\parallel}\left(2c_{1,0,0}+c_{1,1,0}+c_{2,0,0}\right)\right.
\nonumber \\
&&+8\eta_{\parallel}c_{1,0,0}\left(2\cos^{4}\frac
{q_{y}}{2}+2\cos^{2} \frac {\sqrt{3}q_{x}}{2}\cos^{2}\frac
{q_{y}}{2}\right.
-\left.\left.\cos^{2}\frac {\sqrt{3}q_{x}}{2}-3\cos^{2}\frac
{q_{y}}{2}\right)\right)\nonumber\\
&&+J^{2}_{\perp}\left(1-\cos q_{z}\right)\left(1+2\eta_{\perp}\left(c_{0,0,1}+c_{0,0,2}\right)
-4\eta_{\perp}c_{0,0,1}\left(1+\cos q_{z}\right)\right)\nonumber\\
&&+J_{\parallel}J_{\perp}\left(-8\eta_{\perp}c_{1,0,1}\left(1+\cos
q_{z}\right)-2\eta_{\perp}c_{0,0,1}\left(1-\cos q_{z}\right)\right.\nonumber\\
&&-\left.6\eta_{\parallel}c_{1,0,0}\cos q_{z}+22\eta_{\perp}c_{1,0,1}\right)\nonumber\\
&&-\frac
{J_{\parallel}}{2}\left(J_{\parallel}\left(1+2\eta_{\parallel}\left(2c_{1,0,0}+c_{1,1,0}+c_{2,0,0}\right)\right)\right.\nonumber\\
&&+\left.4J_{\perp}\left(\eta_{\perp}c_{0,0,1}\left(1-\cos
q_{z}\right)+\eta_{\perp}c_{1,0,1}-\eta_{\parallel}c_{1,0,0}\cos
q_{z}\right)\right) D_{\bf q},
\label{eq10}
\end{eqnarray}
where $D_{\bf q}$ is defined in Eq. (\ref{dq}).
The correlators $c_{l,k,m}$
have to be determined self-consistently; their indices
correspond to a vector 
${\bf{R}}=l{\bf{a}}_{1}/2+k{\bf{a}}_{2}/2+m{\bf{a}}_{3}$ connecting two spins, 
i.e. 
$c_{l,k,m}=c_{\bf{R}}=\langle S^+_{0} S^-_{{\bf R}}
\rangle= \frac{2}{3}\left\langle {\bf{S}}_{0}{\bf{S}}_{{\bf{R}}}\right\rangle$. 
Using the NN
correlators
$c_{1,0,0}$ and $c_{0,0,1}$ the
GS energy per spin  $E_0/N$ is given by
$
E_0/N=3J_{\parallel}c_{1,0,0}+3J_{\perp}c_{0,0,1}/2.
$
The Eqs. (\ref{eq10a}) - (\ref{eq10}) 
contain eight parameters
$c_{1,0,0}$, $c_{1,1,0}$, $c_{2,0,0}$, $c_{0,0,1}$, $c_{1,0,1}$, $c_{0,0,2}$,
$\eta_{\parallel}$, and $\eta_{\perp}$, which must be determined self-consistently. 
The relation between the correlators and the corresponding Green's 
functions is given by the spectral theorem\cite{tya,elk}. 
Its  application to the correlators 
$c_{l,k,m}$
leads to  
six equations, a seventh  one is the sum rule $c_{0,0,0}=1/2$.
One additional equation is obtained requiring that the matrix of the static
susceptibility $\chi^{+-}_{{\bf{q}}} =
\chi^{+-}_{{\bf{q}}}(\omega=0)$ is isotropic in the limit 
${\bf{q}}\to 0$ \cite{ihle1}, 
i.e. $\lim_{q_x\to 0, q_y \to 0} \chi^{+-}_{{\bf{q}}}|_{q_z=0}
=\lim_{q_z\to 0} \chi^{+-}_{{\bf{q}}}|_{q_x=q_y=0}$.
From that constraint we obtain 
\begin{eqnarray}
&&c_{1,0,0}\left(J_{\perp}\left(1-2\eta_{\perp}\left(3c_{0,0,1}-c_{0,0,2}\right)\right)
+8J_{\parallel}\eta_{\perp}\left(c_{1,0,1}-c_{0,0,1}\right)\right)\nonumber\\
&&-c_{0,0,1}\left(J_{\parallel}\left(1-2\eta_{\parallel}\left(4c_{1,0,0}-c_{1,1,0}-c_{2,0,0}\right)\right)\right.
\left.
+4J_{\perp}\left(\eta_{\perp}c_{1,0,1}-\eta_{\parallel}c_{1,0,0}\right)\right)=0
.
\label{eq12}
\end{eqnarray}
For the discussion of the magnetic GS ordering we consider the 
spin-spin correlation functions 
$\langle {\bf S}_{m\alpha}{\bf S}_{n\beta}\rangle$. 
A possible  magnetic LRO in the system is described via 
$\langle {\bf S}_{m\alpha} {\bf
S}_{n\beta}\rangle$ at large distances ${\bf R}_n - {\bf R}_m$. More
precisely, we consider, as in Refs. \onlinecite{shima,winter}, 
a condensation term
$C_{{\bf{Q}}\alpha\beta}$ in the Fourier transformation $
S_{\alpha \beta}({\bf q})$ of $\langle {\bf S}_{m\alpha} {\bf
S}_{n\beta}\rangle$ according to
\begin{equation}
\langle {\bf S}_{m\alpha} {\bf
S}_{n\beta}\rangle=\frac
{3}{N}\sum_{{\bf{q}}\neq{\bf{Q}}}S_{\alpha\beta}\left({\bf{q}}\right)\exp\left(-i{\bf{q}}{\bf{r}}_{m\alpha,n\beta}\right)+\frac
{3}{2}C_{{\bf{Q}}\alpha\beta}\exp\left(-i{\bf{Q}}{\bf{r}}_{m\alpha,n\beta}\right),
\label{eq14a}
\end{equation}
where {\bf{Q}} is the
corresponding wave vector of the magnetic order and 
$S_{\alpha\beta}\left({\bf{q}}\right)$ is given by
\begin{equation}
S_{\alpha\beta}({\bf{q}})=\frac{3}{2}
\sum_{j=1}^{3}\frac
{m_{j{\bf{q}}}}{2\omega_{j{\bf{q}}}}\left\langle\alpha\right.
\left|j{\bf{q}}\right\rangle\left\langle
j{\bf{q}}\right.\left|\beta\right\rangle.
\label{eq14}
\end{equation}
 The existence of
$C_{{\bf{Q}}\alpha\beta} \ne 0$ is accompanied 
by a diverging static
susceptibility $\chi^{+-}_{{\bf{q}}}$
at the magnetic
wave vector ${\bf{Q}}$.

To describe the magnetic order in a short-range ordered phase we use in
addition to the
spin-spin correlation functions   
the correlation length $\xi_\nu$ along a certain direction ${\bf e}_\nu$.  
To calculate $\xi_\nu$ we apply the procedure
illustrated in Refs. \onlinecite{shima,winter}. 
We expand that eigenvalue of the static susceptibility
$\chi^{+-}_{{\bf{q}}}$, which
is largest at  
the magnetic wave vector ${\bf{Q}}$ around this point ${\bf Q}$. 
The  relevant magnetic 
wave vector ${\bf{Q}}$ in the short-range ordered  
phase is that vector, where the 
largest eigenvalue of the   static susceptibility 
has its maximum.
Then 
the square of the correlation length along ${\bf e}_\nu$
is   given by the factor in front of
the  square of the corresponding component of the ${\bf q}$-vector 
$q_\nu = {\bf e}_\nu
({\bf Q } - {\bf q})$. 
For illustration  we give the expression for the
intralayer correlation length  $\xi_\parallel$ in the pure \kagome limit
($J_\perp=0$), 
$\xi^2_{\parallel}=-2\eta_{\parallel}c_{1,0,0}/
\left(1+2\eta_{\parallel}\left(2c_{1,0,0}+c_{1,1,0}+c_{2,0,0}\right)\right)$,
and omit the lengthy general expressions for the inter- and 
intralayer correlation lengths.  

\section{Results}
\subsection{Antiferromagnetic interlayer coupling $J_{\perp}>0$}
\begin{figure}
\begin{center}
\epsfig{file=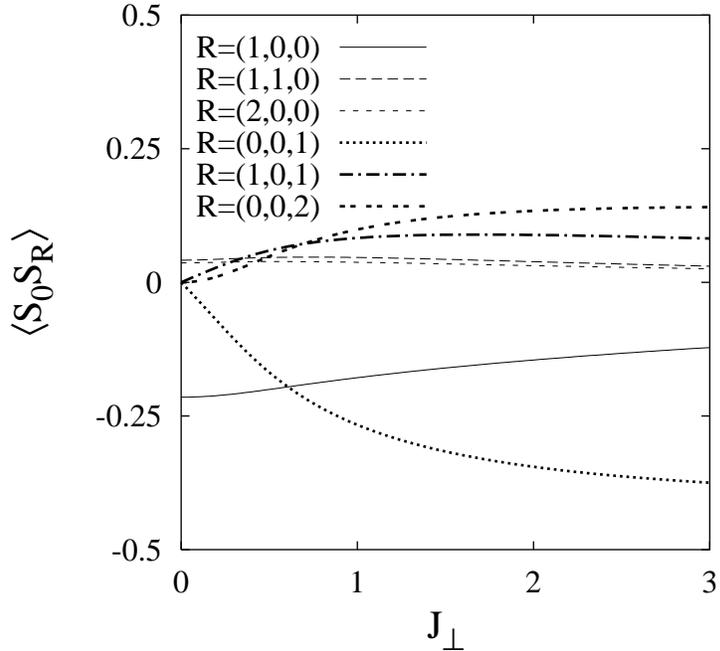,scale=0.50,angle=270.0}
\end{center}
\caption{First, second and third neighbor intralayer 
[${\bf R} =(1,0,0); \;
(1,1,0); \;(2,0,0)$] and interlayer [${\bf R} =(0,0,1); \;
(1,0,1); \;(0,0,2)$] correlation functions
in dependence on the antiferromagnetic
interlayer coupling $J_\perp$.
}
\label{cor_afm}
\end{figure}

Solving the set of the eight self-consistency equations we find that there is
no solution with a non-zero condensation term. Therefore we conclude 
that within the applied 
rotation-invariant Green's function decoupling scheme there
is no indication of magnetic LRO for the stacked 3D \kagome HAFM with
antiferromagnetic interlayer coupling. This remarkable result is
in accordance with the findings within the LSWT (see Sec. \ref{lswt}). 
To study the influence of the interlayer coupling on the 
GS magnetic short range order  we calculate  the spin-spin
correlation functions and the correlation lengths.
For convenience we set $J_{\parallel}=1$.
In 
 Fig.~\ref{cor_afm} we show several first, second and third neighbor
correlation functions.
The strongest change is seen in the strengths 
of the interlayer correlators which increase in the interval $0\le
J_{\perp} \lesssim 1$, for $J_{\perp} \approx 1$ their values are already
close to those of the pure linear chain.  
The change in  the intralayer correlators is small. While the 
NN intralayer correlator
shows some decrease in strength, the second and third neighbor correlators
are weakly increased for small $J_{\perp}$. Remarkably, 
already for $J_{\perp} < 1$
the interlayer correlators become stronger than the intralayer
correlators. 
Note that in the limit $J_{\perp}=0$ our values for the GS energy 
as well as
for the correlators coincide with available results reported in Refs.
\onlinecite{yu,bern}. 

\begin{figure}
\begin{center}
\epsfig{file=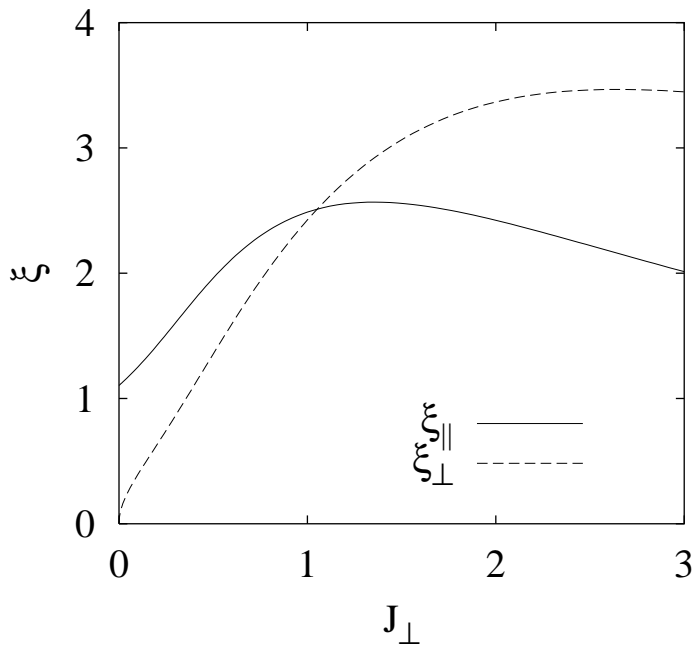,scale=0.50,angle=270.0}
\end{center}
\caption{Intralayer ($\xi_\parallel$) and interlayer ($\xi_\perp$) correlation lengths 
in dependence on the antiferromagnetic
interlayer coupling $J_\perp$.
}
\label{length_afm}
\end{figure}

As illustrated above, to calculate the inter- and intralayer correlation 
lengths $\xi_\perp$ and $\xi_\parallel$ we
have to expand the extreme eigenvalue of the 
static susceptibility around ${\bf Q}$, which is 
${\bf{Q}}=\left(0,0,\pi\right)$ for all $J_\perp >0$. 
The results for the intra- and interlayer
correlation lengths are shown in Fig. \ref{length_afm}.  
The intralayer correlation length 
is of the order of the NN separation $d=1$ in the
pure 2D limit. It increases up to a maximal value of about
$\xi_{\parallel} \approx 2.5 d$ for
$J_{\perp} \approx 1$ and goes to zero for $J_{\perp} \to \infty$. The
increase of the interlayer correlation length  is stronger. 
Though we expect a
diverging     
$\xi_{\perp} $ for the isolated chain, 
it remains finite for any $J_{\perp} > 0$.
This behavior is a drawback of the Green's-function decoupling, 
which is not able to describe the critical
GS with a power-law 
decay of the correlation functions of the linear-chain HAFM.

\subsection{Ferromagnetic interlayer coupling $J_{\perp} < 0$} \label{ferro}
Now we turn to the case $J_{\perp}<0$ which appears to be
more complicated from the
numerical point of view. Starting from $J_{\perp}=0$ we find 
a solution of the
self-consistency equations with  
zero condensation term until $J_{\perp}\approx  - J_{\parallel} = -1$. 
That means
we can conclude that for ferromagnetic interlayer coupling there is no
indication for magnetic LRO for $|J_{\perp}| \lesssim  1$, too. 
Beyond
this point our numerical  procedure fails because of numerical
instabilities in the ${\bf q}$ space integrals.
The reason for that consists in the specific behavior of the dispersion
relations $\omega_{\alpha{\bf{q}}}$ from Eqs. 
(\ref{eq10aa}), (\ref{eq10b}) and (\ref{eq10}).
Increasing the strength of the ferromagnetic interlayer coupling beyond
$|J_{\perp}| \gtrsim 1$ we find that $\omega_{1{\bf{q}}}$ becomes more and
more a flat zero mode in the $q_x$-$q_y$ plane at $q_z=0$, and at the same
time $\omega_{2{\bf{q}}}$ and $\omega_{3{\bf{q}}}$ become more and more
degenerate leading to divergent terms in ${\bf q}$ integrals. 
This behavior of the 
$\omega_{\alpha{\bf{q}}}$  is in coincidence with our findings within the 
LSWT,
where we have found a flat zero-mode and two
degenerate modes in the $q_x-q_y$-plane at $q_{z}=0$ (see
Sec. \ref{lswt}). 
We believe, that the physical interpretation of this behavior
is connected with the classical limit (large quantum number $s$) of the pure
2D \kagome HAFM. In the large $s$ limit the results of the LSWT theory 
become reliable. In the 3D stacked \kagome lattice with ferromagnetic
$J_{\perp}$    
 the
spins along the chains become stronger  coupled 
with increasing $\left|J_{\perp}\right|$ and can be considered as large $s$ 
(quasi-classical) composite spins. These quasi-classical
spins are coupled kagom\'{e}-like with each other, thus 
we are most likely faced with the
classical HAFM on the kagom\'{e} lattice with its huge non-trivial
degeneracy of the GS. Though we do not have rigorous statements
 on 
the zero-temperature correlation functions of the classical \kagome HAFM, 
presumably  at $T=0$ the average of the spin-spin correlation over the set
of all ground states  does not exhibit magnetic LRO.
\cite{moessner01,wir04,chalker92} 
Hence we may conclude that the zero-temperature spin-spin 
correlation functions within a \kagome
layer do not exhibit LRO for any 
ferromagnetic $J_{\perp}$.

However, the subtile
interplay between the tendency to form large $s$ (quasi-classical) composite
spins  and the still remaining (weak) quantum fluctuations may lead to order by
disorder effects. 
As it was argued in Refs. \onlinecite{chalker92,huse92,reimers93,zhito02}
for the pure 2D classical HAFM on the \kagome lattice 
at low temperatures $T \to 0$, 
the entropy of fluctuations may lead to different
relative Boltzmann weights of the classical ground states this way favoring
planar ordered ground states. Presumably there is a nematic LRO but an
algebraic decay of the
spin-spin correlation for $T \to 0$ \cite{chalker92,reimers93}.
Hence,   
for the stacked system considered in this paper at $T=0$  
and for $J_\perp \to
- \infty$, due to 
remaining quantum fluctuations a nematic order
accompanied by 
an algebraic decay in the intralayer spin-spin correlation functions
is possible. 

\begin{figure}
\begin{center}
\epsfig{file=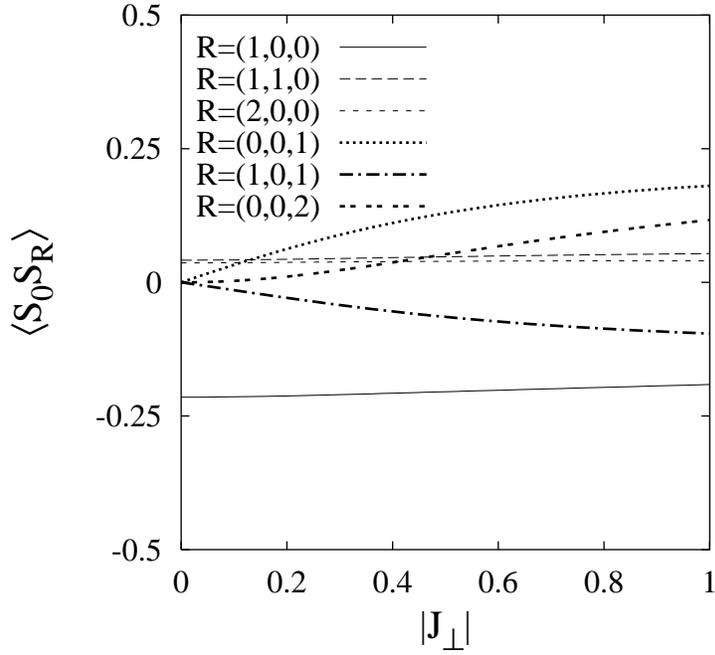,scale=0.50,angle=270.0}
\end{center}
\caption{First, second and third neighbor intralayer [${\bf R} =(1,0,0); \;
(1,1,0); \;(2,0,0)$] and interlayer [${\bf R} =(0,0,1); \;
(1,0,1); \;(0,0,2)$] correlation functions
in dependence on the strength of the ferromagnetic
interlayer coupling $|J_\perp|$.
}
\label{cor_fm}
\end{figure}

\begin{figure}
\begin{center}
\epsfig{file=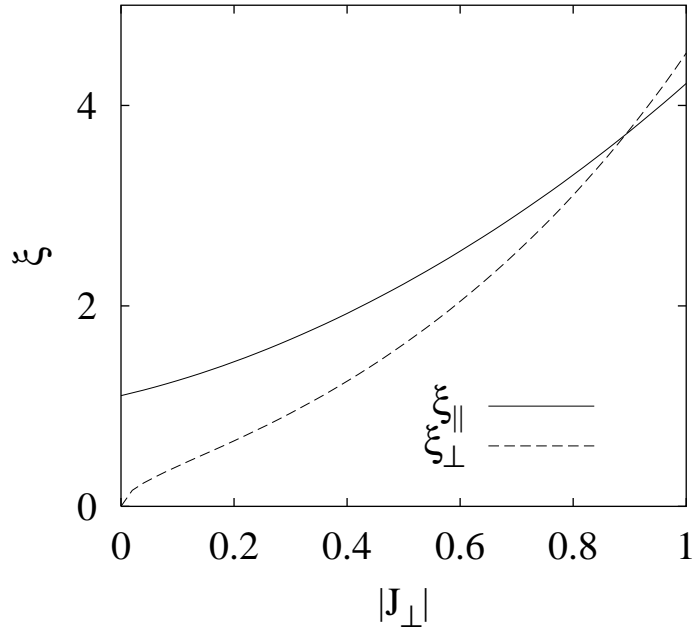,scale=0.50,angle=270.0}
\end{center}
\caption{Intralayer ($\xi_\parallel$) and 
interlayer ($\xi_\perp$) correlation lengths 
in dependence on the strength of the ferromagnetic
interlayer coupling $|J_\perp|$.
}
\label{length_fm}
\end{figure}

To describe the magnetic GS in more detail we present, similarly to the case
$J_\perp > 0$,
 the spin-spin correlation functions 
and the correlation lengths
in Figs. \ref{cor_fm}, \ref{length_fm}. Due to the above
mentioned numerical problems our data are restricted to $0 \le |J_\perp |
\le J_{\parallel} = 1$.   
Several first, second and third neighbor
correlation functions are shown in  
 Fig.\ref{cor_fm}.
Again the change in  the intralayer correlators is very small, whereas 
the strengths of interlayer correlation functions increase with growing 
$|J_{\perp}|$ and become larger than those  of intralayer correlation
functions except
for the NN correlation.

The correlation lengths for $J_\perp < 0$ belong to 
${\bf{Q}}=(0,0,0)$. We show the intra- ($\xi_\parallel$) and 
interlayer ($\xi_\perp$) 
correlation lengths in Fig. \ref{length_fm}. $\xi_\parallel$ and 
$\xi_\perp$ increase in the parameter region shown but are still of the
order of the NN separation $d$. Obviously, the increase in 
$\xi_\perp$ (i.e. along the chains) is significantly stronger. We expect
that 
$\xi_\perp$ diverges for large $|J_\perp |$. As discussed above, 
also a diverging 
$\xi_\parallel$ for $J_{\perp} \to -\infty$  would be possible indicating  
an algebraic decay of the correlation functions.
\section{Summary}
In this paper we have investigated the GS of the stacked three-dimensional 
\kagome spin half Heisenberg antiferromagnet for
ferromagnetic and antiferromagnetic interlayer coupling. To study the
magnetic GS order we have applied a second-order Green's function 
decoupling
scheme going beyond the  RPA. 
Though one could expect that an increase in 
the dimension from two to three would stabilize magnetic order,
we find some evidence that the interlayer coupling is not able to
create magnetic LRO 
 within the antiferromagnetic  \kagome layers. 
These findings based on the Green's function scheme
are supported by linear spin wave theory.

{\it   Acknowledgment:}
This work was supported by the DFG (projects Ri615/12-1 and Ih13/7-1).

\end{document}